\documentclass[manuscript]{aastex62}
\usepackage[T1]{fontenc}

\usepackage{bm}        
\usepackage{amssymb, amsmath}   
\usepackage{cancel}
\usepackage{color}

\newcommand\aastex{AAS\TeX}

\DeclareMathAlphabet{\mathsfit}{\encodingdefault}{\sfdefault}{m}{sl}
\SetMathAlphabet{\mathsfit}{bold}{\encodingdefault}{\sfdefault}{bx}{sl}
\newcommand{\vect}[1]{\bm{#1}}

\DeclareMathOperator{\sech}{sech}
\newcommand{\tensorsym}[1]{\bm{\mathsfit{#1}}}

\received{\today}
\revised{\today}
\accepted{\today}
\submitjournal{ApJ}

\shorttitle{\aastex\ Mass-ratio Dependence in Magnetic Reconnection}
\shortauthors{Li et al.}

\begin{document}

\title{Particle acceleration in kinetic simulations of non-relativistic
magnetic reconnection with different ion-electron mass ratio}

\correspondingauthor{Xiaocan Li}
\email{xiaocanli@lanl.gov}

\author[0000-0001-5278-8029]{Xiaocan Li}
\affil{Los Alamos National Laboratory, Los Alamos, NM 87545, USA}

\author{Fan Guo}
\affiliation{Los Alamos National Laboratory, Los Alamos, NM 87545, USA}

\author{Hui Li}
\affiliation{Los Alamos National Laboratory, Los Alamos, NM 87545, USA}

\begin{abstract}
  By means of fully kinetic particle-in-cell simulations, we study whether the
  proton-to-electron mass ratio $m_i/m_e$ influences the energy spectrum and underlying
  acceleration mechanism during magnetic reconnection. While kinetic simulations are
  essential for studying particle acceleration during magnetic reconnection, a reduced
  $m_i/m_e$ is often used to alleviate the demanding computing resources, which leads to
  artificial scale separation between electron and proton scales. Recent kinetic
  simulations with high-mass-ratio have suggested new regimes of reconnection, as electron
  pressure anisotropy develops in the exhaust region and supports extended current layers.
  In this work, we study whether different $m_i/m_e$ changes the particle acceleration
  processes by performing a series of simulations with different mass ratio
  ($m_i/m_e=25-400$) and guide-field strength in a low-$\beta$ plasma. We find that
  mass ratio does not strongly influence reconnection rate, magnetic energy conversion, ion
  internal energy gain, plasma energization processes, ion energy spectra, and the
  acceleration mechanisms for high-energy ions. Simulations with different mass ratios are
  different in electron acceleration processes, including electron internal
  energy gain, electron energy spectrum and the acceleration efficiencies for high-energy
  electrons. We find that high-energy electron acceleration becomes less efficient when
  the mass ratio gets larger because the \textit{Fermi}-like mechanism associated with particle
  curvature drift becomes less efficient. These results indicate that when particle curvature drift
  dominates high-energy particle acceleration, the further the particle kinetic scales
  are from the magnetic field curvature scales ($\sim d_i$), the weaker the acceleration
  will be.
\end{abstract}


\section{Introduction}
In many solar, space and astrophysical systems, magnetic reconnection is a major mechanism
for energizing plasmas and accelerating nonthermal particles~\citep{Zweibel2009Magnetic}.
A remarkable example is solar flares, where reconnection is observed to trigger efficient
magnetic energy release~\citep{Lin1976Nonthermal}, heats the coronal plasma
\citep[e.g.][]{Caspi2010RHESSI, Longcope2010Model}, and accelerates both electrons and ions
into nonthermal power-law energy distributions~\citep{Krucker2010Measure, Krucker2014Particle,
Oka2013Kappa, Oka2015Electron, Shih2009RHESSI}. However, how particles are accelerated over
a large-scale reconnection region is still not well understood.

The dynamics of magnetic reconnection is believed to involve both the macroscopic scales
($>10^6$ m in solar flares) and the kinetic scales ($< 10$ m in solar flares)
\citep{Daughton2009Transition, Ji2011Phase}, and thus a multi-scale approach is essential
for understanding particle acceleration during reconnection. Starting from the kinetic scales, kinetic
simulations (fully kinetic or hybrid) are often used to study how particles are accelerated
and coupled with background fluids~\citep[e.g.][]{Drake2006Electron}. Various models are then
developed to capture these processes for studying the macroscopic particle acceleration
\citep{Drake2018Comp, LeRoux2015Kinetic, LeRoux2016Combining, LeRoux2018Self, Li2018Large,
Montag2017Impact, Zank2014Particle, Zank2015Diffusive} and are applied in explaining
local particle acceleration between interacting flux ropes in the solar wind~\citep{Zhao2018Unusual,
Zhao2019Particle, Adhikari2019Role}. Previous kinetic simulations have
identified that the reconnection X-line region~\citep{Hoshino2001Suprathermal, Drake2005Production,
Fu2006Process, Oka2010Electron, Egedal2012Large, Egedal2015Double, Wang2016Mechanisms} and
contracting and merging magnetic islands~\citep{Drake2006Electron, Oka2010Electron, Liu2011Particle,
Drake2013Power, Nalewajko2015Distribution} are the major particle acceleration sites during
reconnection. Under the guiding-center approximation, recent simulations have also identified
particle curvature drift motion along the motional electric field as the major particle
acceleration mechanism~\citep{Dahlin2014Mechanisms, Guo2014Formation, Guo2015Particle,
Li2015Nonthermally, Li2017Particle}.~\citet{Li2018Roles} further showed that
the flow compression and shear effects well capture the primary particle energization,
as in the standard energetic particle transport theory~\citep{Parker1965Passage, Zank2014Transport,
LeRoux2015Kinetic}. Fluid compression and shear effects have also been used to
quantify plasma energization during island coalescence problem~\citep{Du2018Plasma}.
The connection between particle acceleration associated with particle drift motion
and that related to fluid motion is summarized in Appendix of current paper. \citet{Li2018Roles}
also found that flow compression and shear are suppressed as the guide field increases.
To alleviate the computational cost, these previous simulations were mostly carried out
using a reduced proton-to-electron mass ratio $m_i/m_e=25$.

A higher mass ratio ($m_i/m_e\geqslant400$), however, can potentially change the plasma
energization and particle acceleration processes, because different magnetic field,
currents, and pressure anisotropy structures emerge as $m_i/m_e$ becomes
larger~\citep[e.g.][]{Egedal2013Review, Egedal2015Double, Le2013Regimes}. By performing
kinetic simulations of reconnection with different mass ratio, guide field, and plasma
$\beta$,~\citet{Le2013Regimes} demonstrated that the magnetic fields and currents fall
into four regimes, and that the transition guide field between different regimes changes
with the mass ratio and plasma $\beta$. They also identified a new regime with an extended
current layer only when $m_i/m_e\geqslant400$. Those works were mostly focused on
the dynamics and structures in the reconnection layer. Therefore, it is worthwhile to
understand how the mass ratio influences the plasma energization and particle acceleration
processes.

In this paper, we focus on the consequences of having a disparity between the energy
releasing scale (the radius of magnetic curvature $\sim$ the ion inertial length $d_i$)
and the plasma kinetic scales (the electron gyroradius $\rho_e$ to $d_i$). The scale
separation between electrons and protons becomes larger as the mass ratio approaches the
realistic value. For example, $\rho_e/d_i=\sqrt{\beta_i}\sqrt{T_e/T_i}\sqrt{m_e/m_i}$
decreases with $m_i/m_e$, where the ion plasma $\beta_i$ and the temperature ratio $T_e/T_i$
are usually fixed.

Here we perform fully kinetic particle-in-cell simulations with $m_i/m_e=25$, 100, and 400
to study whether the mass ratio changes the plasma energization and particle acceleration
processes during magnetic reconnection. For each mass ratio, we perform four runs with
different guide field: 0, 0.2, 0.4, and 0.8 times of the reconnection magnetic field,
so the series of simulations covers all the regimes studied by~\citet{Le2013Regimes}. In Section
\ref{sec:methods}, we describe the simulation parameters. In Section~\ref{sec:res},
we present the results on how the energy conversion, reconnection rate, particle energy spectra,
plasma energization processes, and particle acceleration rates change with the mass ratio and
the guide field strength. In Section~\ref{sec:con}, we discuss the conclusions and the
implications based on our simulation results.

\section{Numerical Simulations}
\label{sec:methods}
\begin{figure*}[ht!]
  \centering
  \includegraphics[width=0.5\textwidth]{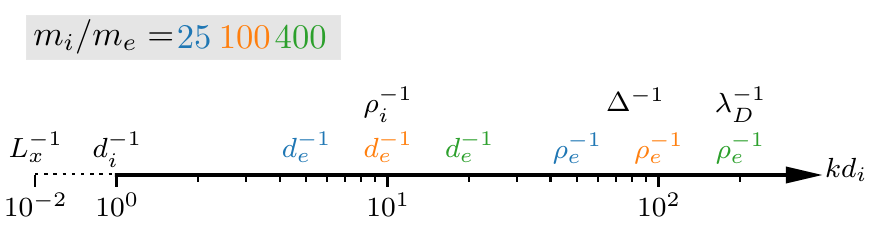}
  \caption{
    \label{fig:spatial_scales}
    The spatial scales normalized by the ion inertial length $d_i$ in the simulations
    with different mass ratio $m_i/m_e$. $d_e=c/\omega_{pe}$ is the electron inertial
    length. $\rho_e=v_\text{the}/\Omega_{ce}$ is the electron gyroradius.
    $\rho_i=v_\text{thi}/\Omega_{ci}$ is the ion gyroradius, which is the
    same in terms of $d_i$ for different $m_i/m_e$. $\lambda_D$ is the Debye length,
    which is the same for different $m_i/m_e$. $\Delta$ is the simulation cell size.
    We also include $L_x^{-1}$, which is $0.01d_i^{-1}$, in the plot.
  }
\end{figure*}

We carry out 2D kinetic simulations using the VPIC particle-in-cell code~\citep{Bowers2008PoP},
which solves Maxwell's equations and the relativistic Vlasov equation.
The simulations start from a force-free current sheet with $\vect{B}=B_0\tanh(z/\lambda)\hat{x} +
B_0\sqrt{\sech^2(z/\lambda) + b_g^2}\hat{y}$, where $B_0$ is the strength of the
reconnecting magnetic field, $b_g$ is the strength of the guide field $B_g$ normalized by $B_0$,
and $\lambda$ is the half-thickness of the current sheet. Note that for this paper we will use $B_g$
and $b_g$ interchangeably when it does not cause confusion. We preform simulations with $B_g=0.0$,
0.2, 0.4, and 0.8 in three mass ratios: 25, 100, and 400. All simulations have the same
Alfven speed $v_\text{A}$ ($=B_0/\sqrt{4\pi n_0m_i}$) and electron beta $\beta_e=8\pi nkT_e/B_0^2$
defined using reconnecting component of the magnetic field. We choose $\lambda=d_i$ for all
simulations, where $d_i=c/\omega_\text{pi}=c/\sqrt{4\pi n_ie^2/m_i}$ is the ion inertial
length. The initial particle distributions are Maxwellian with uniform density $n_0$ and
temperature $T_i=T_e=T_0$. The temperature is taken to be $kT_0=6.25\times 10^{-4}m_ec^2$,
$0.0025m_ec^2$, and $0.01m_ec^2$ for runs with $m_i/m_e=25$, 100, and 400, respectively,
where $m_ec^2$ is fixed for runs with different mass ratio. Electrons are set to have a bulk
velocity drift $U_e$ so the Ampere's law is satisfied. The ratio of electron plasma frequency and
electron gyrofrequency $\omega_{pe}/\Omega_{ce}=4$, 2, and 1 for runs with $m_i/m_e=25$, 100, and 400,
respectively. The resulting Alfv\'en speed is $0.05c$ and the electron beta is 0.02 for all simulations.
The domain sizes are $L_x\times L_z=100d_i\times 50d_i$, and the grid sizes are $8192\times4096$
for all simulations. Figure~\ref{fig:spatial_scales} shows that the electron kinetic scales
($\rho_e$ and $d_e$) deviate more from the energy releasing scale ($\sim d_i$) as the mass ratio
becomes larger. We use 400 particles per cell per species in all simulations. As the mass ratio
increases, both the plasma skin depth and gyroradius are at scales shorter than one ion skin
depth. For electric and magnetic fields, we employ periodic boundaries along the $x$-direction
and perfectly conducting boundaries along the $z$-direction. For particles, we employ periodic
boundaries along the $x$-direction and reflecting boundaries along the $z$-direction.
Initially, a long wavelength perturbation with $B_z=0.02B_0$ is added to induce
reconnection~\citep{Birn2001Geospace}. The simulations are terminated around $t\Omega_{ci}=100$
(one Alfv\'en crossing time) to minimize the effect of the periodic boundary conditions
along the $x$-direction.

\section{Results}
\label{sec:res}
\subsection{Current Layer Structures}
\begin{figure*}[ht!]
  \centering
  \includegraphics[width=\textwidth]{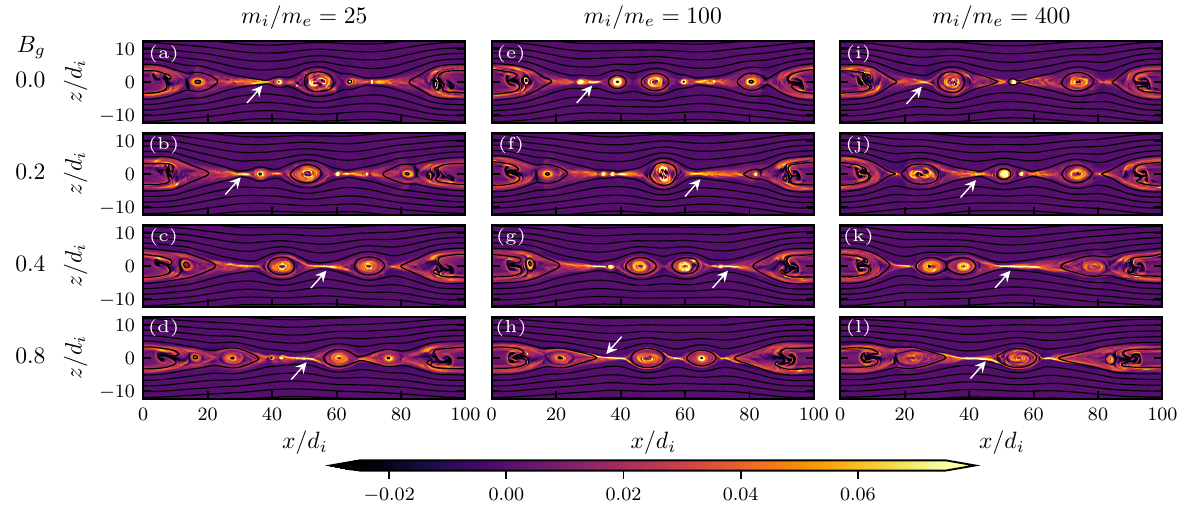}
  \caption{
    \label{fig:jy_2d}
    The out-of-plane current density $j_y$ at $t\Omega_{ci}=80$, 83, and 86 for
    $m_i/m_e=25$, 100, and 400, respectively. We choose different time frames because
    the reconnection onset is faster in the runs with a lower mass ratio
    (see Figure~\ref{fig:rrate}). The unit of $j_y$ is $en_0c$. The white arrows
    point out regions to be discussed in the main text.
  }
\end{figure*}

As the simulations proceed, current layers are unstable to the tearing instability,
leading to fractional sheets filled with magnetic islands. Figure~\ref{fig:jy_2d}
shows the out-of-plane current density $j_y$ for runs with three mass ratios $25 - 400$
with different guide fields from $B_g = 0$ to $0.8$. The time steps shown are
$t\Omega_{ci}=80$, 83, and 86 for $m_i/m_e=25$, 100, and 400, respectively. We choose slightly
different time frames because reconnection onsets slightly faster in the runs with a lower
mass ratio. Overall, the current layers vary in length and are oriented along different directions
depending on the guide-field strength. In the low guide-field regime, an elongated current
layer emerges because of an unmagnetized electron jet formed in the electron diffusion region
(panels (b), (e), (f), and (i)). Since there is a finite $B_y$ field in the center of a
force-free current sheet even when $B_g=0$, electrons could be magnetized in the low
guide-field regime, and localized current layers are formed instead (panels (a) and (j)).
A new regime, first studied by~\citet{Le2013Regimes}, emerges with extended current layers
embedded in the reconnection exhaust when $m_i/m_e\geqslant 400$ (panel (k)). These current
layers can extend over $20d_i$ and therefore might affect particle energization processes.
In contrast, the current layers are shorter in runs with a lower mass ratio (panels (c)
and (d)). As the guide field gets even stronger (panels (d), (h), and (l)), the electrons
become well magnetized, and the current density tends to peak at one of the diagonal branches
of the reconnection separatrix.~\citet{Le2013Regimes} studied the scaling extensively and
found that these structures are regulated by the electron pressure anisotropy and the
properties of the electron orbits, which depend on the mass ratio and guide field.
The scaling in our simulations does not exactly match with the diagram by~\citet{Le2013Regimes}
(Figure 3 in their paper). This is because the force-free current sheet (different from
the Harris current sheet used in~\citet{Le2013Regimes}) has a finite magnetic field
along the guide-field direction in the center of the current sheet even when $B_g=0$,
and also because these structures are dynamic and can be destroyed as the simulations evolve.
In summary, different mass ratio results in different types of current layers, especially
when $B_g\leqslant0.4$. In the following sections we will study whether the
mass-ratio dependence influences the mechanisms for plasma energization and particle
acceleration processes.

\subsection{Reconnection Rate}
\begin{figure*}[ht!]
  \centering
  \includegraphics[width=\textwidth]{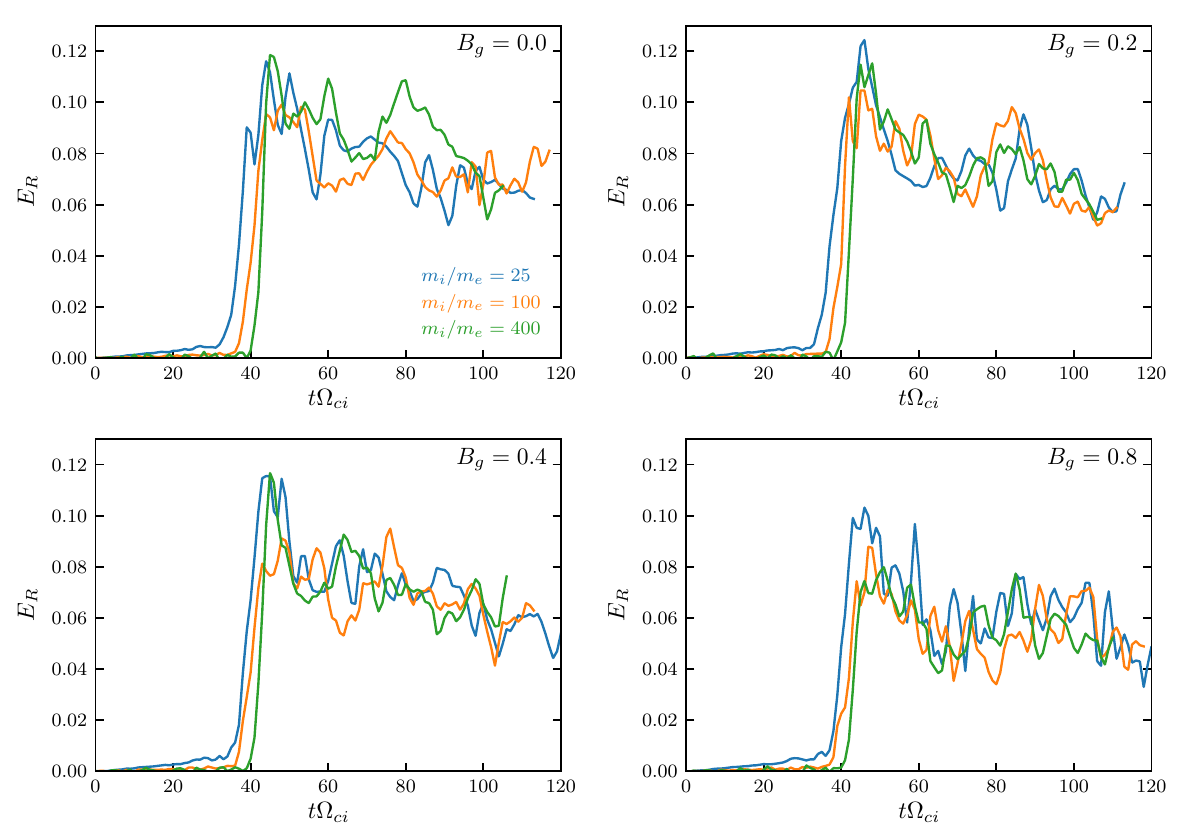}
  \caption{
    \label{fig:rrate}
    Normalized reconnection rate $E_R=(\partial\psi/\partial t)/(B_0v_\text{A})$
    \citep{Daughton2009Transition}, where $\psi=\max(A_y) - \min(A_y)$ along $z=0$,
    $A_y$ is the $y$ component of the vector potential, $B_0$ is the strength of
    the initial reconnecting component of the magnetic field, and
    $v_\text{A} \equiv B_0/\sqrt{4\pi n_0 m_i}$ is the Alfv\'en speed defined by $B_0$ and the
    initial particle number density $n_0$.
  }
\end{figure*}

Before diving into the energization processes, we check whether the different mass ratio
changes the reconnection rate. Following~\citet{Daughton2009Transition}, we evaluate
the normalized reconnection rate $E_R\equiv(\partial\psi/\partial t)/(B_0v_\text{A})$,
where $\psi=\max(A_y) - \min(A_y)$ along $z=0$, $A_y$ is the $y$ component of the
vector potential, and $v_\text{A} \equiv B_0/\sqrt{4\pi n_0 m_i}$ is the Alfv\'en
speed defined by $B_0$ and the initial particle number density $n_0$. Figure~\ref{fig:rrate}
shows that the reconnection rate for various cases. Since we do not average the rate over
a long time interval~\citep{Daughton2009Transition}, the rate fluctuates rapidly.
Figure~\ref{fig:rrate} shows that the reconnection onset is faster in the runs with
a lower mass ratio. In the following analysis, unless specified otherwise, we will
shift the runs with $m_i/m_e=100$ by $-3\Omega_{ci}^{-1}$ and the runs with $m_i/m_e=400$
by $-6\Omega_{ci}^{-1}$ to match the reconnection onset. Figure~\ref{fig:rrate} shows that
the reconnection rate is roughly the same for runs with different mass ratio. The peak
reconnection rate is about 0.1 for runs with $B_g\leq 0.4$, consistent with previous
kinetic simulations~\citep[e.g.][]{Birn2001Geospace}. The peak rate does not sustain,
because the periodic boundary conditions limit the simulation durations~\citep{Daughton2006Fully},
and because we use the upstream plasma parameters ($B_0$ and $n_0$) instead of that
near the dominant reconnection $x$ point \citep{Daughton2007Collisionless}. At
$t\Omega_{ci}=100$ (one Alfv\'en crossing time), $E_R$ decreases to about $0.06$.
We will terminate our analysis at $t\Omega_{ci}=100$, when only a few large islands and smaller
secondary islands are left in the simulations.

The evolution of reconnection rate  shows that the runs are similar in the reconnected magnetic fluxes.
Previous kinetic simulations have shown that the converted magnetic energy can be channelled
into plasma kinetic energy preferentially by the parallel electric field $\vect{E}_\parallel$
near the reconnection $X$-line and by the \textit{Fermi}-like mechanism associated with
contracting and merging magnetic islands~\citep[e.g.][]{ Dahlin2014Mechanisms, Guo2014Formation,
Li2015Nonthermally, Li2017Particle}. $\vect{E}_\parallel$ accelerates particles proportionally
to their velocities; the \textit{Fermi}-like mechanisms accelerate particles
proportionally to their energies. The dominant mechanism could be different for particles
with different energies and for electrons and ions.
The mass ratio could change the relative importance of these mechanisms, leading
to different particle energy distributions and energy partition between electrons and ions.
The following analysis will show how the mass ratio changes the plasma energization and
particle acceleration processes.

\subsection{Energy Conversion}
\begin{figure*}[ht!]
  \centering
  \includegraphics[width=\textwidth]{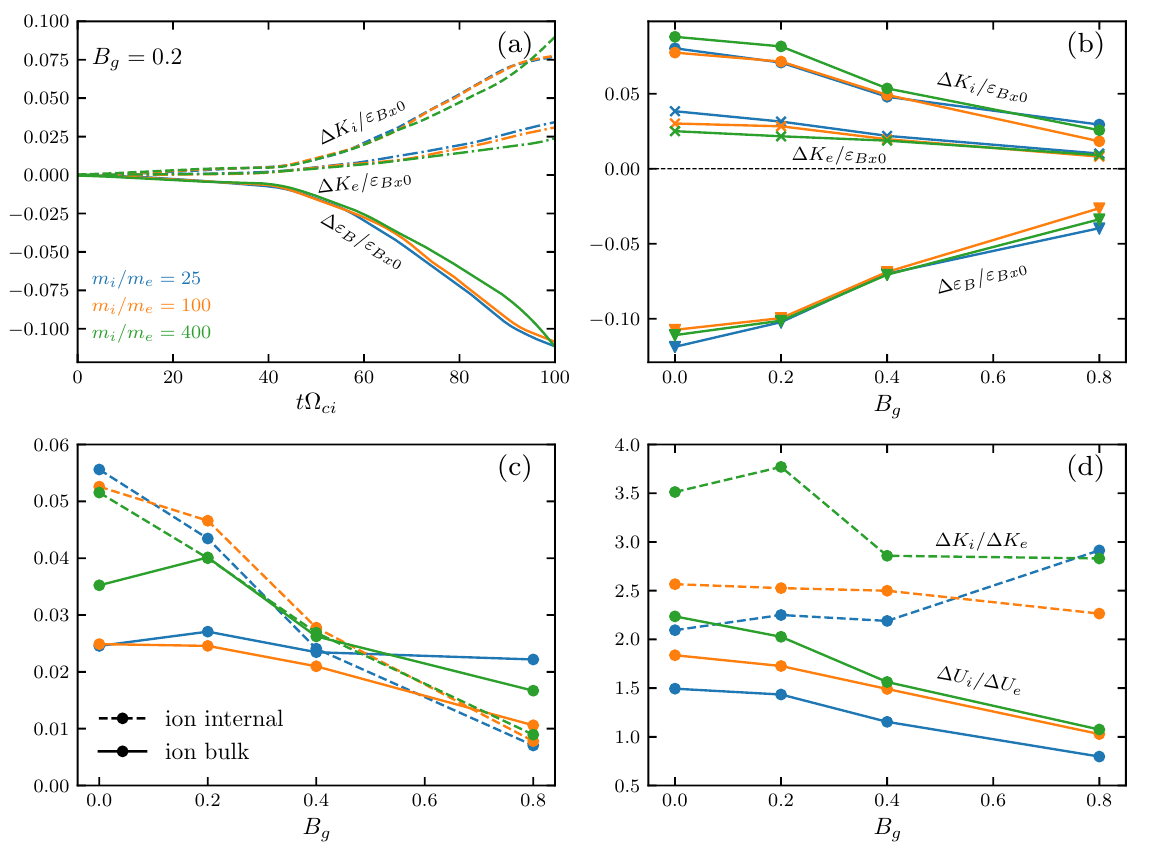}
  \caption{
    \label{fig:ecov}
    Energy conversion for different runs.
    (a) Time evolution of the change of the magnetic energy $\Delta\varepsilon_B$
    (solid), the electron energy gain $\Delta K_e$ (dash-dotted), and the ion
    energy gain $\Delta K_i$ (dashed) for the runs with $B_g=0.2$. The energies
    are normalized by the initial energy of the reconnecting component of the magnetic
    field $\varepsilon_{Bx0}$. We have shifted the runs with $m_e/m_i=100$ by
    $-3\Omega_{ci}^{-1}$ and the runs with $m_i/m_e=400$ by $-6\Omega_{ci}^{-1}$
    as described in Figure~\ref{fig:rrate}.
    (b) The changes of the magnetic energy (triangle), the electron energy gain
    (star), and the ion energy gain $\Delta U_i$ (circle) accumulated to $t\Omega_{ci}=100$.
    (c) Ion internal energy gain and bulk energy gain. They are also normalized
    by $\varepsilon_{Bx0}$. The internal energy density is calculated from the diagonal
    components of the ion pressure tensor as $\sum_i P_{ii}/2$. The ion bulk energy
    density is $0.5n_im_iu_i^2$, where $u_i$ is the ion bulk flow speed.
    (d) The energy partition between ions and electrons. The dashed lines are for the
    total kinetic energies; the solid lines are for the internal energies.
  }
\end{figure*}

We start investigating the energization processes by examining the energy evolution
in the simulations. Figure~\ref{fig:ecov} shows the energy conversion in these simulations
till $t\Omega_{ci}=100$. Panel (a) shows the time evolution of the change of the magnetic
energy $\Delta\varepsilon_B$, the electron energy gain $\Delta K_e$, and the ion energy
gain $\Delta K_i$ in the runs with $B_g=0.2$ . We normalize them by the initial energy
of the $x$ component (reconnecting component) of the magnetic field
$\varepsilon_{Bx0} = B_x^2/(8 \pi)$. Similar fraction of magnetic energy
(11\% of $\varepsilon_{Bx0}$) is
converted into plasma kinetic energy in all runs. Panel (b) shows that slightly more magnetic
energy is converted in runs with $m_i/m_e=25$ and 400 when $B_g=0.0$ or 0.8, and similar
fraction of magnetic energy is converted for the other cases. Panel (a) shows that as
$m_i/m_e$ gets larger, electrons gain less energy, resulting in about 31\%, 28\%, and 21\%
of $\Delta\varepsilon_{B}$ going into electrons in the runs with $m_i/m_e=25$, 100, and 400,
respectively. Panel (b) shows that the difference gets smaller as $B_g$ increases. When $B_g=0.8$,
electrons gain a similar fraction of converted magnetic energy in runs with different mass ratio.
Panel (a) also shows that ions gain less energy first and then more energy to the end of the
simulation with $m_i/m_e=400$. Panel (b) shows that ions do gain more energy in runs with
$m_i/m_e=400$ than the other runs, except when $B_g=0.8$, ions gain most energy in the run
with $m_i/m_e=25$. The guide-field dependence of different energies shown in panel (b)
is consistent for different mass ratio despite the differences in their actual values.

Since the reconnection outflow is about the Alfv\'en speed $0.05c$, the ion bulk energy
is significant in our simulations. Panel (c) shows that, depending on the guide field,
the ion bulk energy is comparable with or even larger than the ion internal energy.
Panel (c) also shows that the ion bulk energy is larger in the runs with $m_i/m_e=400$
except when $B_g=0.8$, and that it does not change much with the guide field when
$m_i/m_e=25$, while it generally gets weaker as $m_i/m_e$ becomes larger. In contrast,
the ion internal energy always decreases as $B_g$ gets larger, and the difference between
different mass ratio is subtle. As a result, $\Delta K_i/\Delta K_e$ does not show clear
dependence on the guide field, while $\Delta U_i/\Delta U_e$ decreases as $B_g$ becomes
larger (panel (d)). When $B_g=0.8$, $\Delta U_i/\Delta U_e$ approaches one for the cases
with $m_i/m_e=100$ or 400 and becomes even smaller in the run with $m_i/m_e=25$.
Panel (d) also shows that $\Delta U_i/\Delta U_e$ is much larger in runs with a higher
mass ratio, especially in the low guide-field cases. We expect $\Delta U_i/\Delta U_e$
will be even larger in simulations with a real $m_i/m_e=1836$. In summary, a lower mass
ratio helps reconnection to convert more magnetic energy into electron kinetic energy and
a similar amount of magnetic energy into ion internal energy, which changes the energy
partition between electrons and ions. Then, the next question is whether a different mass
ratio results in different electron distributions but similar ion distributions, which
we now discuss.

\subsection{Particle Energy Distributions}
\begin{figure*}[ht!]
  \centering
  \includegraphics[width=\textwidth]{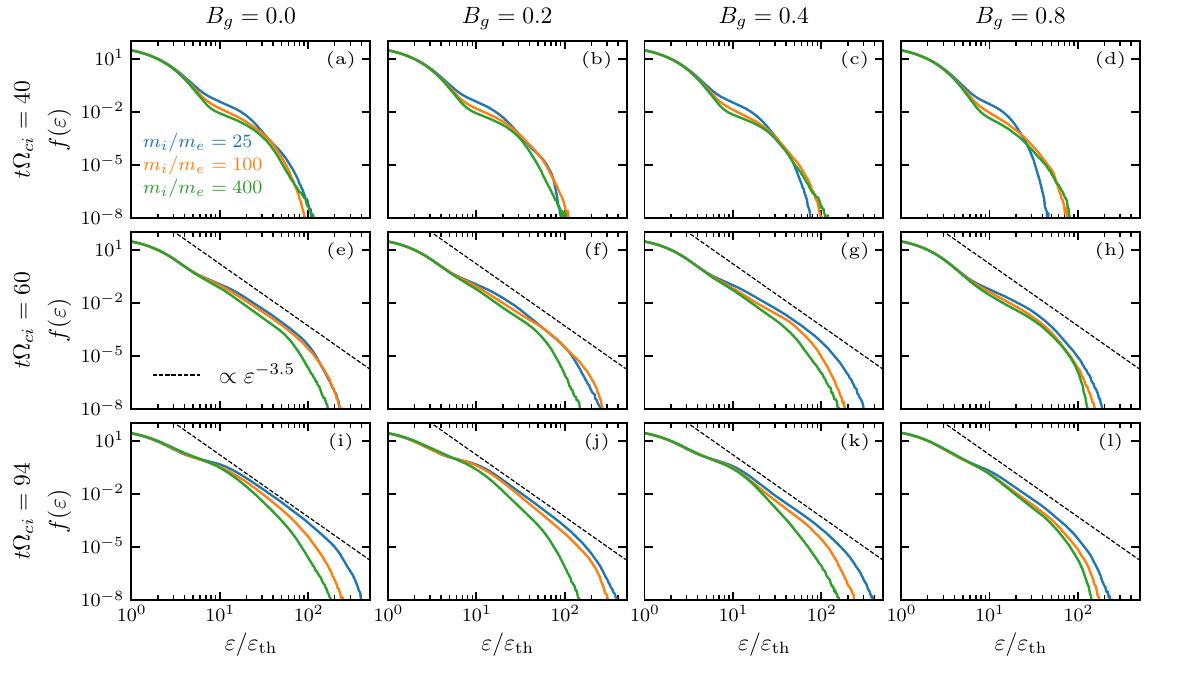}
  \caption{
    \label{fig:espect}
    Normalized electron energy spectra
    $f(\varepsilon)=(dN(\varepsilon)/d\varepsilon)/(N_\text{tot}
    \varepsilon_\text{th}^{400}/\varepsilon_\text{th})$ at $t\Omega_{ci}=40$,
    60, and 94, where $N_\text{tot}\equiv n_x\times n_z\times\text{nppc}$ is the total
    number of macro electrons in the simulation, $\varepsilon_\text{th}=3kT_0/2$ is the initial
    thermal energy for different mass ratio, and $\varepsilon_\text{th}^{400}$ is the
    initial thermal energy for $m_i/m_e=400$. The electron kinetic energy
    $\varepsilon \equiv (\gamma-1)m_ec^2$ is normalized by $\varepsilon_\text{th}$,
    where $\gamma$ is the Lorentz factor. Note that we shifted the runs with $m_i/m_e=100$
    by $-3\Omega_{ci}^{-1}$ and the runs with $m_i/m_e=400$ by $-6\Omega_{ci}^{-1}$ to
    match the reconnection onset. The dashed lines indicate power-law distributions with
    a power index $-3.5$. Note that they are not fitted distributions but only a guide
    for the analysis.
  }
\end{figure*}

Figure~\ref{fig:espect} shows the normalized electron energy spectra for all electrons
at $t\Omega_{ci}=40$, 60, and 94. Electrons are accelerated to over 100 times of the
initial thermal energy $\varepsilon_\text{th}$ in all runs. The accelerated
electrons develop a significant high-energy tail ($>10\varepsilon_\text{th}$), which
contains 0.7--4\% of all electrons and 7--38\% of the total electron kinetic energy
to the end of the simulations ($t\Omega_{ci}=100$). Top panels show that electrons
quickly reach $100\varepsilon_\text{th}$, and that the acceleration is faster in the runs
with $m_i/m_e=100$ or 400. As studied by previous kinetic simulations, parallel electric field
$\vect{E}_\parallel$ plays a key role in the acceleration, for that $E_\parallel$ not
only accelerates most electrons near the reconnection X-line~\citep{Li2017Particle,
Lu2018Formation} but also forms pseudo electric potential wells, which trap electrons
so that they can be further accelerated by perpendicular electric field $\vect{E}_\perp$
\citep{Egedal2015Double}. As a result, most electrons near the X-line are accelerated to
develop flat spectra that appear to be hard power-law distributions for
$\varepsilon\in[20, 50]\varepsilon_\text{th}$~\citep{Li2015Nonthermally}. But these spectra
are usually transient, because they only contain less than 10\% of the high-energy electrons
($>10\varepsilon_\text{th}$) at $t\Omega_{ci}=100$, and because these electrons are trapped
near the center of the primary magnetic islands~\citep{Li2017Particle}. As the simulations
evolve to $t\Omega_{ci}=60$ (panels (e)--(h)), the electron acceleration in the runs with
$m_i/m_e=25$ catches up and becomes the strongest especially in runs with $B_g=0.4$ or 0.8
(panels (g) and (h)).  The spectra appear to be power-law distributions with a power index
-3.5 (dashed lines) for $\varepsilon\in[10, 100]\varepsilon_\text{th}$, especially in the
runs with $B_g=0$ and 0.2 (panels (e) and (f)). But these spectra are actually the
superposition of a series of thermal-like distributions in different sectors of a 2D
magnetic island~\citep{Li2017Particle}. To the end of the simulations (panels (i)--(l)),
the separation between different mass ratio becomes even larger. The spectra in the runs
with $m_i/m_e=25$ still appear to be power-laws with as power index -3.5, and the spectra
are much steeper in the runs with higher mass ratios. These results indicate that a lower
proton-to-electron mass ratio tends to overestimate the high-energy electron acceleration.

\begin{figure*}[ht!]
  \centering
  \includegraphics[width=\textwidth]{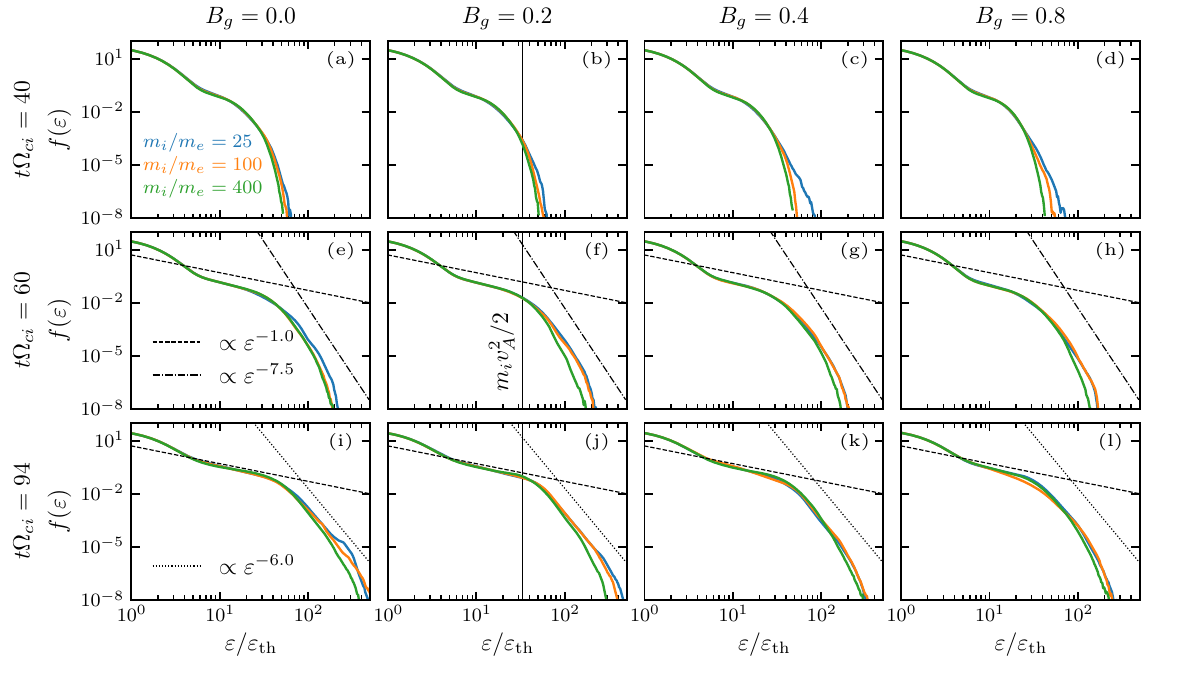}
  \caption{
    \label{fig:ispect}
    Normalized ion energy spectra
    $f(\varepsilon)=(dN(\varepsilon)/d\varepsilon)/(N_\text{tot}
    \varepsilon_\text{th}^{400}/\varepsilon_\text{th})$ at $t\Omega_{ci}=40$,
    60, and 94, where $N_\text{tot}\equiv n_x\times n_z\times\text{nppc}$ is the total
    number of macro ions in the simulation, $\varepsilon_\text{th}=3kT_0/2$ is the initial
    thermal energy for different mass ratio, and $\varepsilon_\text{th}^{400}$ is the
    initial thermal energy for $m_i/m_e=400$. The ions kinetic energy
    $\varepsilon \equiv (\gamma-1)m_ic^2$ is normalized by $\varepsilon_\text{th}$,
    where $\gamma$ is the Lorentz factor. Note that we shifted the runs with $m_i/m_e=100$
    by $-3\Omega_{ci}^{-1}$ and the runs with $m_i/m_e=400$ by $-6\Omega_{ci}^{-1}$ to
    match the reconnection onset.
    The dashed lines indicate power-law distributions with a power index $-1.0$.
    The dotted lines indicate power-law distributions with a power index $-6.0$.
    The dash-dotted lines indicate power-law distributions with a power index $-7.5$.
    Note that they are not fitted distributions but only a guide for the analysis.
    The vertical black lines indicate the bulk kinetic energy of a single ion advected
    by the reconnection outflow ($\approx v_\text{A}$).
  }
\end{figure*}

Figure~\ref{fig:ispect} shows the normalized ion energy spectra for all ions at $t\Omega_{ci}=40$,
60, and 94. Ions are accelerated up to $500\varepsilon_\text{th}$, higher than electrons.
The accelerated ions develop significant high-energy tails. At the beginning ($t\Omega_{ci}=40$),
ions are quickly accelerated to the reconnection outflow speed $\approx v_\text{A}$.
This process does not increase the ion internal energy much but rather accelerates all ions
in the reconnection exhausts to a bulk kinetic energy of $0.5m_iv_\text{A}^2$. We find that
the acceleration is associated with particle polarization drift instead of the parallel electric
field as for electrons (more discussion in Figure~\ref{fig:pene_i}).  As the simulations
evolve to $t\Omega_{ci}=60$ (panels (e)--(h)), the spectra in the runs
with $m_i/m_e=25$ and 100 are close to each other, and the fluxes of high-energy ions
in the runs with $m_i/m_e=400$ are still lower. The spectra appear to be power-laws
$\propto\varepsilon^{-1}$ for $\varepsilon$ around $10\varepsilon_\text{th}$. The high-energy
tail is likely a drift Maxwellian distribution with a drift energy $\approx 0.5m_iv_\text{A}^2$,
because the break point of the spectra is about $0.5m_iv_\text{A}^2$ (vertical solid
line in panel (f)). To the end of the simulations (panels (i)--(l)), the low-energy
part is still a hard power low $\propto \varepsilon^{-1}$, and the high-energy tail becomes
harder and resembles a power-law $\propto\varepsilon^{-6}$. The spectra in the runs with
$m_i/m_e=400$ are still steeper when $B_g=0.0$ or 0.2, but the difference is obvious only
at the highest energies ($\varepsilon>200\varepsilon_\text{th}$). The spectra in
the runs with $B_g=0.4$ or 0.8 are close to each other. We find that high-energy particles
($>0.5m_iv_\text{A}^2$) are further accelerated by the \textit{Fermi}-like mechanism
associated with particle curvature drift (more discussion later). The maximum ion energy
keeps increasing because of the \textit{Fermi}-like mechanism but is limited by the simulation
duration ($100\Omega_{ci}^{-1}\approx$ 16 ion gyro-period). We expect that ions can be accelerated
to higher energies and develop an even harder high-energy tail in larger simulations.
In summary, ions develop similar energy spectra for different mass ratio, and the spectra
have a hard low-energy part and a steep high-energy part, separating by the reconnection
bulk flow energy $0.5m_iv_\text{A}^2$.

\subsection{Plasma Energization}
\begin{figure*}[ht!]
  \centering
  \includegraphics[width=\textwidth]{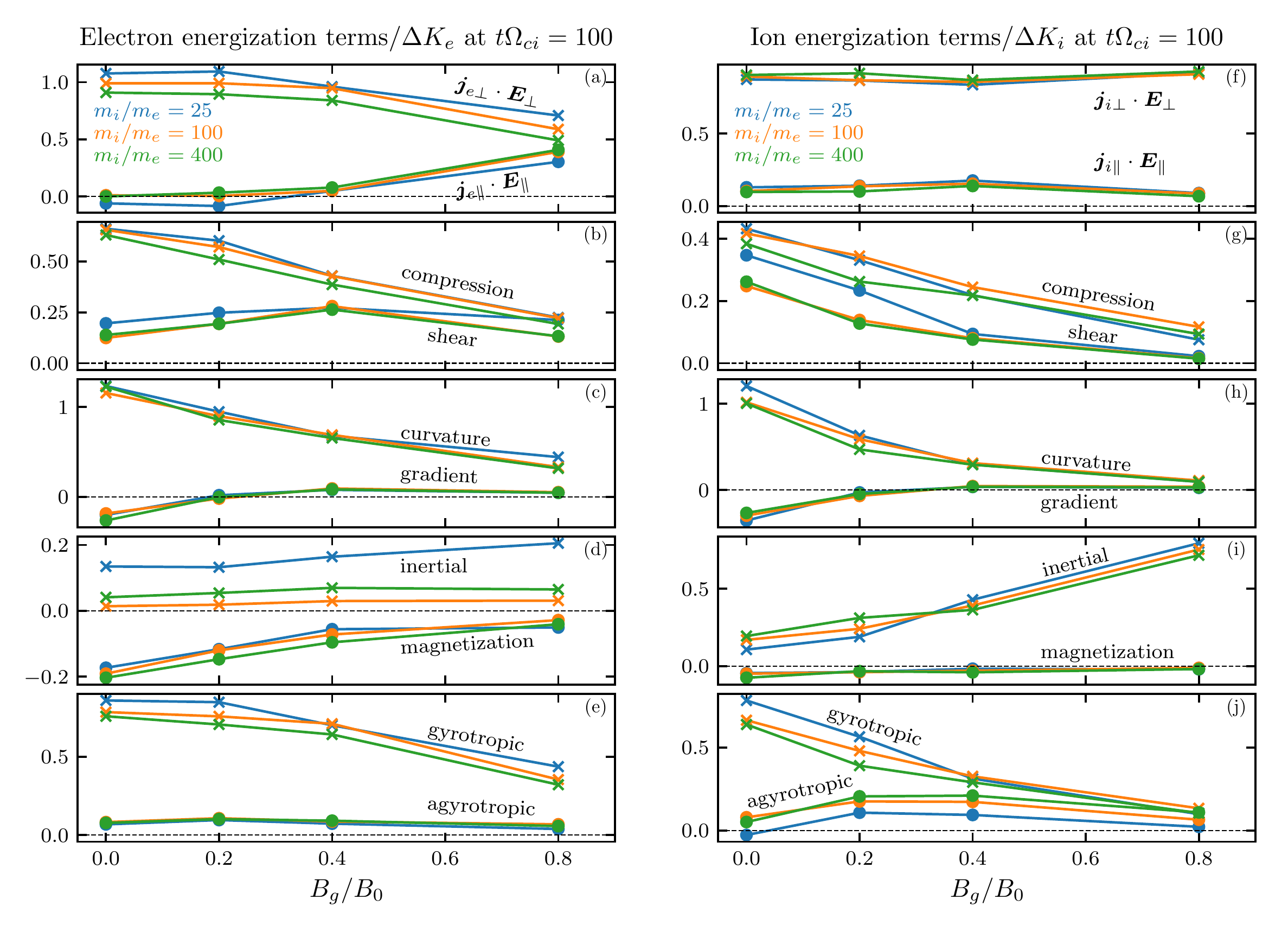}
  \caption{
    \label{fig:fene_e}
    Fluid energization terms accumulated to $t\Omega_{ci}=100$ for electrons
    (left panels) and ions (right panels). All these terms are integrated over the
    whole simulation domain and normalized by the particle energy gain
    ($\Delta K_e$ or $\Delta K_i$) at $t\Omega_{ci}=100$.
    (a) \& (f) Energization by parallel or perpendicular electric field.
    (b) \& (g) Energization associated with flow compression or flow shear
    (see~\ref{equ:comp_shear} for their definitions).
    (c) \& (h) Energization associated with curvature drift or gradient drift
    (see~\ref{equ:jperp_drift} for their definitions).
    (d) \& (i) Energization associated with flow inertia or magnetization
    (see~\ref{equ:jperp_drift} for their definitions).
    (e) \& (j) Energization associated with the gyrotropic pressure tensor
    $(\nabla\cdot\tensorsym{P}_{gs})\cdot\vect{v}_E$ or the agyrotropic pressure tensor
    $(\nabla\cdot\tensorsym{P}_{s} - \nabla\cdot\tensorsym{P}_{gs})\cdot\vect{v}_E$,
    where $\tensorsym{P}_s$ is the whole  pressure tensor for a single species,
    $\tensorsym{P}_{gs} \equiv p_{s\perp}\tensorsym{I} +
    (p_{s\parallel} - p_{s\perp})\vect{b}\vect{b}$ is gyrotropic pressure tensor,
    $p_{s\parallel}$ is the parallel pressure, $p_{s\perp}$ is the perpendicular
    pressure, $\tensorsym{I}$ is the unit dyadic, $\vect{b}$ is the unit vector
    along the local magnetic field direction, and $\vect{v}_E$ is the
    $\vect{E}\times\vect{B}$ drift velocity. Note that the accumulation over time
    could introduce errors since we only have 100 time frames.
  }
\end{figure*}

Plasma energization analysis based on the guiding-center drift description has been
routinely carried out in kinetic simulations for studying particle acceleration mechanisms
\citep{Dahlin2014Mechanisms, Li2015Nonthermally, Li2017Particle, Li2018Roles,
Wang2016Mechanisms}. Figure~\ref{fig:fene_e} shows multiple plasma energization terms
associated with the parallel or perpendicular electric field, flow compression or flow
shear (see~\ref{equ:comp_shear} for their definitions), curvature drift or gradient drift
(see~\ref{equ:jperp_drift} for their definitions), flow inertia or magnetization
(see~\ref{equ:jperp_drift} for their definitions), and gyrotropic or agyrotropic pressure
tensors. For electrons, a mass ratio of 25 tends to overestimates the contributions by
$\boldsymbol{E}_\perp$ (panel (a)), flow compression and shear (panel (b)), flow inertia
(panel (d)), and gyrotropic pressure tensor (panel (e)), but the guide-field dependence
is consistent across runs with different mass ratio. Among these terms, the inertia
term is mostly overestimated in the runs with $m_i/m_e=25$, but it only contributes to
the bulk energization. For ions, Figure~\ref{fig:fene_e} (g) shows that $m_i/m_e=25$ tends
to overestimate the contribution by flow shear when $B_g\leq0.2$, and that $m_i/m_e=25$ or 100
tends to overestimate the contribution by flow compression when $B_g\leq0.2$;
Figure~\ref{fig:fene_e} (j) shows that ions are more gyrotropic in the runs with
$m_i/m_e=25$ than that in the runs with a higher mass ratio. This is because ions become
less well-magnetized when its gyroradius
$\rho_i/d_e=v_\text{thi}\omega_{pe}/(c\Omega_{ci})= \sqrt{m_i/m_e}(v_\text{thi}/v_A)$
gets larger with the mass ratio, where $v_\text{thi}/v_A=\sqrt{\beta_i}$ is the same for
all runs.

Since other energization terms were more or less studied before, we summarize the
results shown in Figure~\ref{fig:fene_e} without going into details. For electrons,
panel (a) shows that most energization is done by $\vect{E}_\perp$ when $B_g<0.5$, and
that the energization by $\vect{E}_\parallel$ becomes comparable with that by
$\vect{E}_\perp$ when $B_g=0.8$; panel (b) shows that flow compression energization
dominates flow shear energization ($\propto$ pressure anisotropy), although the former
keeps decreasing with the guide field, and the latter slightly increases until $B_g=0.4$
because of an increasing pressure anisotropy~\citep{Li2018Roles}; panels (c) and (d)
show that the energization associated with curvature drift dominates the other energization
terms by $\vect{E}_\perp$, and that the energization associated with flow inertia
contributes significantly only when $m_i/m_e=25$; panels (e) shows that the energization
associated with the gyrotropic pressure tensor always dominates the energization associated
with the agyrotropic pressure tensor, indicating that most electrons are well-magnetized
in the simulations. For ions, panel (f) shows that most energization is done by $\vect{E}_\perp$,
and that this does not change much with the guide field; panel (g) shows that compression
energization always dominates shear energization, and that both terms gradually decrease
with the guide field, which is different from that for electrons; panels (h)--(j) show
that the energization associated with curvature drift and flow inertia are two most important
terms for ions besides the energization associated with the agyrotropic pressure tensor,
and that curvature drift dominates when $B_g<0.4$ and flow inertia dominates when
$B_g\geqslant 0.4$. In summary, plasma energization is similar in runs with different
mass ratio, so a lower mass ratio (e.g. 25) is still useful for studying particle
acceleration mechanisms and their scaling with the guide field.

\subsection{Particle Acceleration Rates}
\begin{figure*}[ht!]
  \centering
  \includegraphics[width=0.5\textwidth]{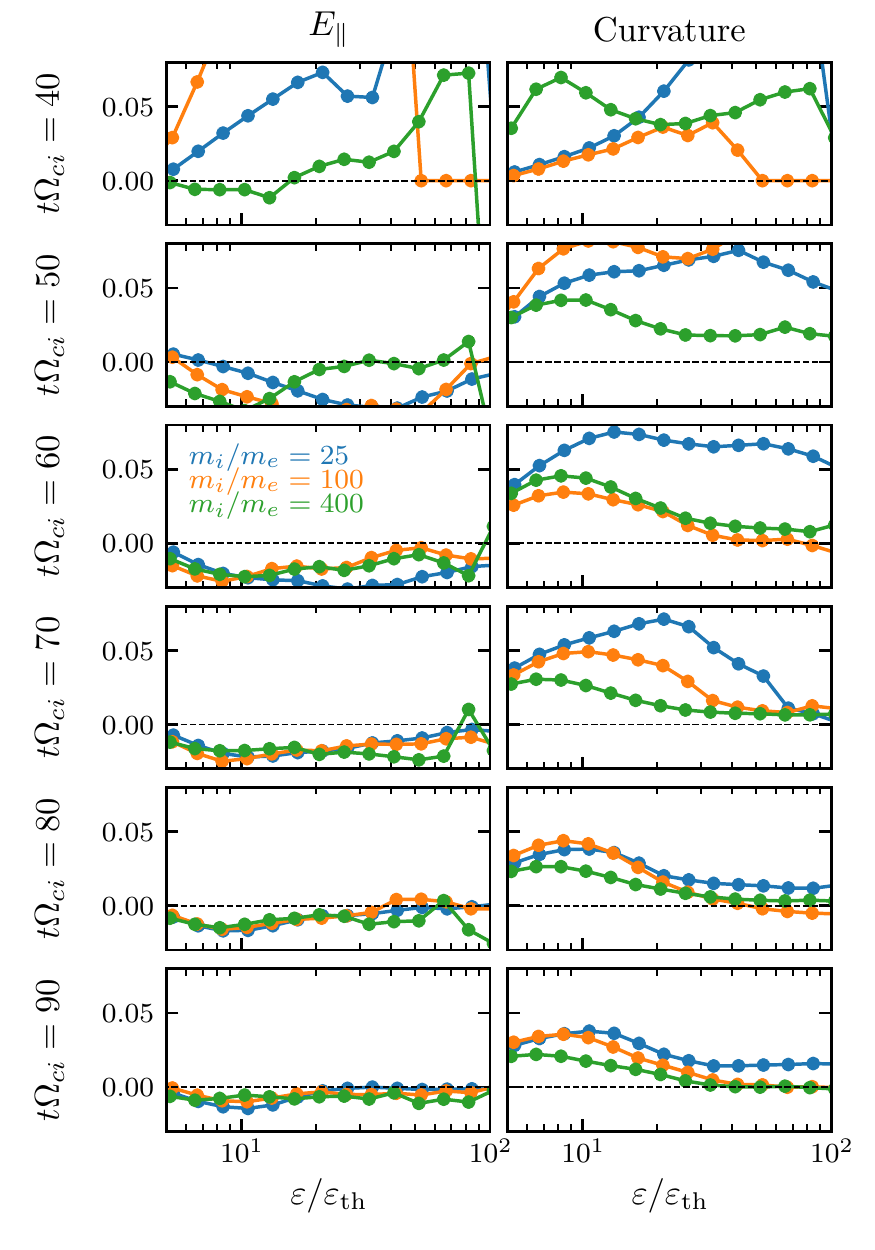}
  \caption{
    \label{fig:pene_e}
    Electron acceleration rate
    $\alpha(\varepsilon, t)\equiv\left<\dot{\varepsilon}(\varepsilon, t)/\varepsilon(t)\right>$
    associated with the parallel electric field, and particle curvature drift for the
    runs with $B_g=0.2$, where $\left<\dots\right>$ is the average for particles in different
    energy bins. We normalize $\alpha$ by $\Omega_{ci}$ to compare among the runs with
    different mass ratio. Since we only have 10 time frames of particle data, we only
    shifted the run with $m_i/m_e=400$ by $-10\Omega_{ci}^{-1}$.
  }
\end{figure*}

To further reveal the difference between runs with different mass ratio, we use all particles
to evaluate the particle acceleration rates
$\alpha(\varepsilon, t)\equiv\left<\dot{\varepsilon}(\varepsilon, t)/\varepsilon(t)\right>$
associated with $\vect{E}_\parallel$, $\vect{E}_\perp$, curvature drift, gradient drift,
parallel drift, inertial drift, polarization drift, and betatron acceleration.
Figure~\ref{fig:pene_e} shows the two largest terms for electrons: $\vect{E}_\parallel$
and curvature drift, for the runs with $B_g=0.2$. Since the simulation duration is
$100\Omega_{ci}^{-1}$ for all runs, in order to compare among the runs with different
mass ratio, we normalize $\alpha$ by $\Omega_{ci}$. We find that $\vect{E}_\parallel$
is efficient at accelerating electrons early in the simulation ($t\Omega_{ci}=40$),
but it does not accelerate or even decelerates energetic electrons ($>10\varepsilon_\text{th}$)
latter. The right panels of Figure~\ref{fig:pene_e} show particle curvature drift
generally leads to acceleration. It gradually decreases as the simulation evolves and
approaches zero for high-energy electrons ($>30\varepsilon_\text{th}$) in the runs with
$m_i/m_e=100$ or 400 but stays finite in the run with $m_i/m_e=25$. Combining the negative
acceleration rate due to $\vect{E}_\parallel$, we find that high-energy electrons are
decelerated latter in the runs with $m_i/m_e=100$ or 400. In contrast, the high-energy
electrons are continuously accelerated in the run with $m_i/m_e=25$, so the ``power-law''
can survive, as show in Figure~\ref{fig:espect}. Note that these results still hold for
runs with a different guide field that are not shown here. In summary, as the mass ratio gets
larger, high-energy electron acceleration becomes less efficient, because the acceleration
rate by $\vect{E}_\parallel$ becomes negative, and because the \textit{Fermi}-like
mechanism becomes less efficient.

\begin{figure*}[ht!]
  \centering
  \includegraphics[width=0.75\textwidth]{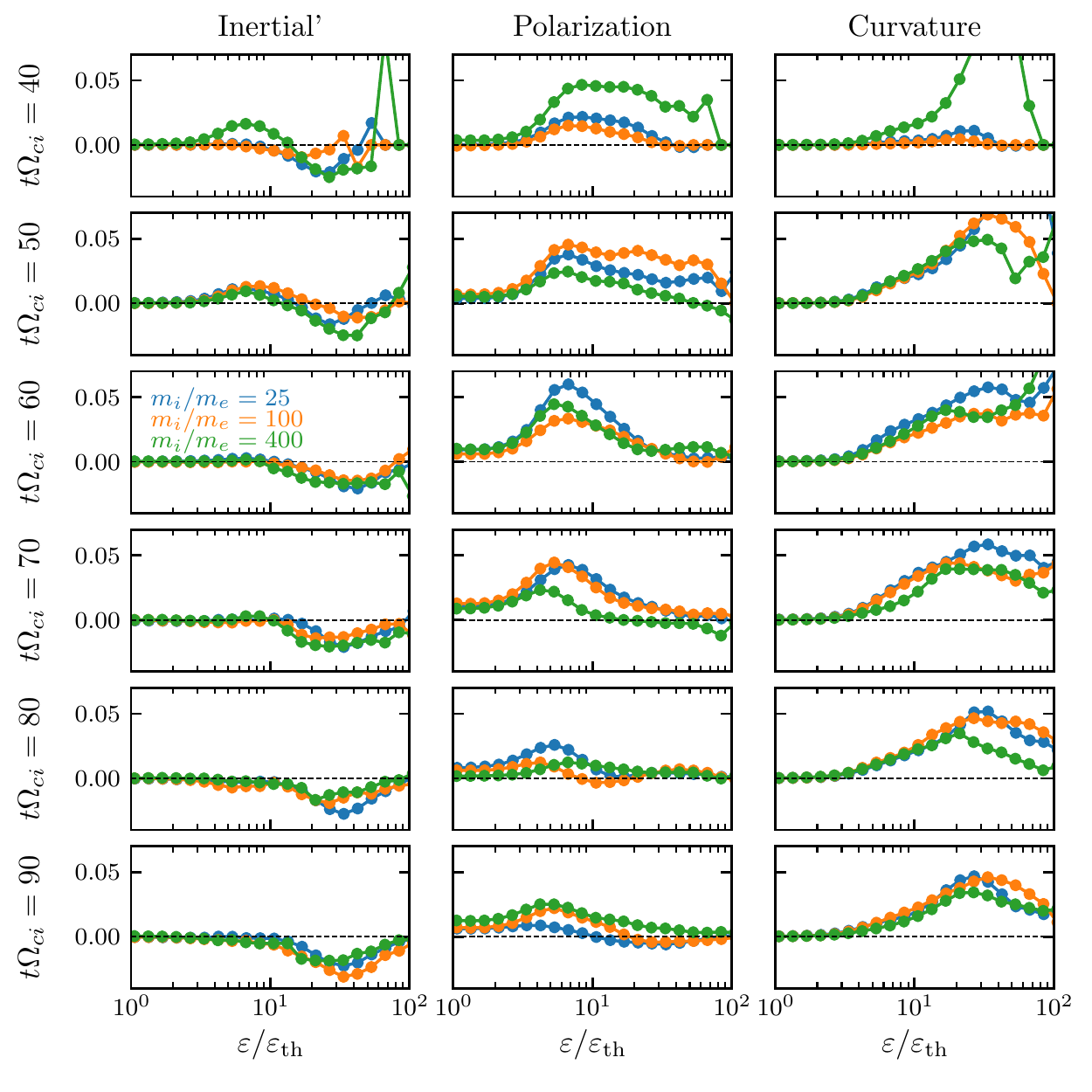}
  \caption{
    \label{fig:pene_i}
    Ion acceleration rate
    $\alpha(\varepsilon, t)\equiv\left<\dot{\varepsilon}(\varepsilon, t)/\varepsilon(t)\right>$
    associated with particle inertial' drift $\equiv$ particle inertial drift - particle
    curvature drift, particle curvature drift, and particle polarization drift for the runs with
    $B_g=0.2$, where $\left<\dots\right>$ is the average for particles in different
    energy bins. We normalize $\alpha$ by $\Omega_{ci}$ to compare among the runs with
    different mass ratio. Since we only have 10 time frames of particle data, we only
    shifted the run with $m_i/m_e=400$ by $-10\Omega_{ci}^{-1}$.
  }
\end{figure*}

Figure~\ref{fig:pene_i} shows the acceleration rates for ions. We find that the acceleration
rates associated with particle inertial drift, polarization drift, and curvature drift are
most important for ions. Since the inertial drift contains particle curvature drift, we
subtract the curvature drift from the inertial drift and call the residue the inertial' drift
in the left panels. The acceleration rate associated with the inertial' drift is negative
for energetic ions with tens of $\varepsilon_\text{th}$, indicating that the acceleration process
associated with the inertial' drift decelerates ions. The middle panels of Figure~\ref{fig:pene_i}
show that $\alpha$ associated with polarization drift is efficiently at accelerating ions
at different energies early in the simulations but peaks around $5\varepsilon_\text{th}$
and approaches zero when $\varepsilon>20\varepsilon_\text{th}$ latter in the simulations.
This indicates that particle polarization drift along $\vect{E}_\perp$ is efficient at
driving the reconnection bulk flow. In contrast,
the right panels of Figure~\ref{fig:pene_i} show that the \textit{Fermi}-like mechanism
associated with particle curvature drift preferentially accelerates ions at high energies
($>20\varepsilon_\text{th}$), and that it is still strong to the end of the simulations.
We expect that ions can be accelerated to higher energies and develop an even harder
high-energy spectra in larger simulations. The right panels show that the acceleration
associated with curvature drift is slightly smaller in the run with $m_i/m_e=400$ than
that in the runs with lower mass ratios. This explains why the high-energy ion fluxes
are lower in the runs with $m_i/m_e=400$, as shown in Figure~\ref{fig:ispect}. These
results on the ion acceleration rates are consistent among the runs with different mass
ratio, suggesting that we could use a lower mass ratio (e.g. 25 or 100) to study ion
acceleration in low-$\beta$ reconnection.

\section{Conclusions and Discussions}
\label{sec:con}
In this work, we study whether and how the proton-to-electron mass ratio affects the
particle acceleration processes in kinetic simulations of magnetic reconnection through
performing simulations with different mass ratio and guide-field strength.
The simulations show different current layer structures that depend on the mass ratio
and guide-field strength, consistent with earlier studies~\citep[e.g.][]{Le2013Regimes}.
We find that simulations with different mass ratios are similar in reconnection rate, magnetic
energy conversion, ion internal energy gain, plasma energization processes, ion energy spectra,
and the acceleration mechanisms for high-energy ions, but simulations show different electron
internal energy gain, electron energy spectrum, and the acceleration efficiencies for high-energy
electrons. We find that electrons gain more energy (internal or kinetic) in runs with
a lower mass ratio. As a result, the ion-to-electron energy partition increases with the
mass ratio, e.g. from 1.5 for $m_i/m_e=25$ to 2.25 for $m_i/m_e=400$ when $B_g=0$. We find that
the electron spectrum gets steeper as the mass ratio gets larger. By calculating the particle
acceleration rates due to different particle guiding-center drift motions, we find that as the
mass ratio increases, high-energy electron acceleration becomes less efficient because parallel
electric field tends to decelerate high-energy electrons, and because the \textit{Fermi}-like
mechanism associated with particle curvature drift becomes less efficient.

The simulations also show that the total plasma energization associated with the guiding-center
drift motions and flow compression and shear is similar for the runs with different mass ratio.
A lower mass ratio tends to overestimate some of the energization terms, but the guide-field
dependence is consistent across runs with different mass ratio. By subtracting the gyrotropic
pressure tensor from the whole pressure tensor, we find that most electrons are well magnetized
even when $B_g=0$, and that the agyrotropic ion distributions contribute over 15\%
of the total ion energization when $m_i/m_e=400$ and $B_g\geqslant 0.2$. This indicates
that ions are not well-magnetized when $m_i/m_e$ is large. These results suggest a
lower mass ratio is still good for studying energy conversion mechanisms during magnetic
reconnection.

The ion acceleration rates show that the acceleration terms associated with ion
inertial drift, polarization drift, and curvature drift are most important for ions.
Ion inertial drift (with curvature drift being subtracted) decelerates high-energy
ions ($>20$ times of the initial thermal energy). Ion polarization drift tends to
drive the reconnection bulk flow and is mostly efficient for low-energy ions (around
5 times of the initial thermal energy). We find that high-energy ions are accelerated
by the \textit{Fermi}-like mechanism associated with particle curvature drift along the
motional electric field.

The ion energy distributions show that ions are accelerated to form Alfv\'enic reconnection
outflow when they enter the reconnection layer. The thermalisation processes (e.g. compression
and shear) result in a much hotter plasma than the inflow plasma. Similar processes could
occur in solar flares, where $v_\text{A}\approx 10^3$ km/s and the ion thermal speed
$v_\text{thi}\approx 10^2$ km/s in the lower solar corona. As indicated by observations
\citep[e.g.][]{Liu2013Plasmoid}, the coronal plasma can be heated from 1 million Kelvin
(MK) to tens of MK in a flare region. The superposition of such multicomponent super-hot
plasmas can even produce the observed coronal hard X-ray emission, as predicted in simulations
by~\citet{Cheung2018Comprehensive}.

We have carried another set of simulations, in which we fix the electron thermal velocity
and $\omega_{pe}/\Omega_{ce}$ (effectively varying the Alfv\'en speed for different mass ratio).
This is typical when using a lower mass ratio to save the computationally costs. We find
that the above conclusions still hold for this new set of simulations.
The consistency between the two sets of simulations suggests once the scale separation
between electrons and ions are fixed, the acceleration processes of a single species
will be similar. Below is our explanation of similar ion acceleration and different electron
acceleration in runs with different mass ratio. The particle acceleration rates
(Figure~\ref{fig:pene_e} and~\ref{fig:pene_i}) show particle curvature drift as the dominant
high-energy acceleration mechanism, and the curvature drift acceleration is most efficient
in the reconnection exhaust ($\sim d_i$). We can treat the $d_i$ as the energy-containing
scale. The closer is particle gyromotion scale to $d_i$, the stronger high-energy acceleration
do we expect. For ions, $\rho_i/d_i$ (0.1 in our simulations) is larger than $\rho_e/d_i$,
so ions tend to be accelerated to higher energies than electrons; $\rho_i/d_i$ is constant
for different mass ratio, so the ion spectra are similar for different mass ratio.
For electrons, $\rho_e/d_i$ gets smaller as the mass ratio increases, so high-energy electron
acceleration gets weaker when $m_i/m_e$ gets larger.
\begin{figure*}[ht!]
  \centering
  \includegraphics[width=0.5\textwidth]{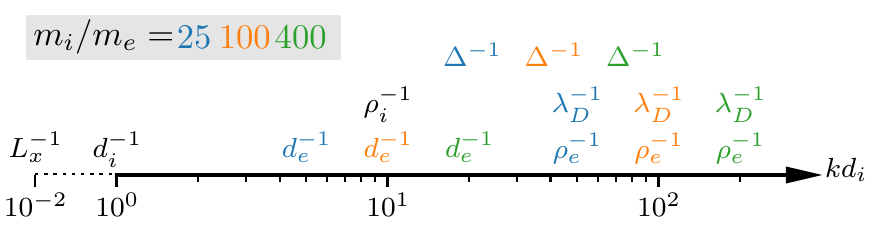}
  \caption{
    \label{fig:spatial_scales2}
    Similar as Figure~\ref{fig:spatial_scales} but for a different set of simulations,
    where we fix the electron thermal speed and $\omega_{pe}/\Omega_{ce}$. We use
    a lower resolution for runs with a lower $m_i/m_e$ because $\lambda_D$ is larger.
  }
\end{figure*}

Although we present results only for runs with a guide field up to $0.8B_0$, we have performed
simulations with a stronger guide field (1.6, 3.2, and $6.4B_0$), which are more relevant to
particle acceleration due to quasi-2D turbulence or interacting small-scale flux ropes in the
inner heliosphere~\citep{Smith2006Turbulent, Zank2014Transport, LeRoux2015Kinetic, Hu2018Automated}.
We find that ion acceleration is still similar for runs with different mass ratio, and
electron acceleration is still less efficient as the mass ratio gets larger.
As the guide field becomes larger than $B_0$, the parallel electric field becomes the dominant
energization mechanism for electrons, but it is inefficient at accelerating energetic electrons,
resulting in a much lower high-energy electron fluxes; the acceleration associated polarization
drift becomes the dominate energization mechanism for ions, but it drives ion bulk flow instead
of accelerating high-energy ions. For both electrons and ions, the acceleration rate associated
with particle curvature drift becomes lower as the guide field gets stronger, indicating that the
acceleration time scale becomes longer. To fully evaluate the effect of curvature drift in the
strong guide-field reconnection, we need much larger simulations that runs for a much longer time.
We defer these studies to a future work.

Our simulations have a few limitations. First, we perform simulations only in low-$\beta$
plasmas with the same temperature for electrons and ions, which are suitable for studying
particle acceleration at the reconnection site of a solar flare. Conclusions on the
mass-ratio dependence might change at the reconnection sites in Earth's magnetosphere or the
accretion disk corona, where ions can be much hotter than electrons, and the plasma $\beta$
can be larger than 0.1. Second, the simulation duration is limited by the box sizes and the
periodic boundary conditions. A larger simulation with more realistic open boundary
conditions could change the relative importance of the acceleration near the
reconnection X-line and the acceleration associated with magnetic islands. Third,
high-energy particles are confined in the 2D magnetic islands and cannot be further
accelerated. The self-generated turbulence in 3D reconnection could change the acceleration
processes and their dependence on the mass ratio.

To conclude, we find that different mass ratios are similar in reconnection rate,
magnetic energy conversion, ion internal energy gain, plasma energization processes,
ion energy spectra, and the acceleration mechanisms for high-energy ions. We find that ion
acceleration is similar for different mass ratio because the dominant acceleration
mechanism for energetic ions is due to particle curvature drift, and it does not change much
with the mass ratio. Runs with different mass ratios are different in electron internal energy
gain, electron energy spectrum, and the acceleration efficiencies for high-energy electrons.
We find that high-energy electron acceleration becomes less efficient when the mass ratio
gets larger because parallel electric field tends to decelerate high-energy electrons,
and because the \textit{Fermi}-like mechanism associated with particle curvature drift
becomes less efficient. These results indicate that when particle curvature drift
dominates high-energy particle acceleration, the further the particle kinetic scale
are from the magnetic field curvature scales ($\sim d_i$), the weaker the acceleration
will be, at least in 2D.

\appendix
\section{Fluid description of plasma energization}
\citet{Li2018Roles} described the plasma energization processes in term of
$\vect{j}_s\cdot\vect{E}$, where the perpendicular component of the current density
 $\vect{j}_s$ for any species is
\begin{align}
  \vect{j}_{s\perp} = p_{s\parallel}\frac{\vect{B}\times(\vect{B}\cdot\nabla)\vect{B}}{B^4} +
  p_{s\perp}\frac{\vect{B}\times\nabla B}{B^3} -
  \left[\nabla\times\frac{p_{s\perp}\vect{B}}{B^2}\right]_\perp +
  \rho_s\frac{\vect{E}\times\vect{B}}{B^2} - n_sm_s\frac{d\vect{u}_s}{dt}\times
  \frac{\vect{B}}{B^2} \label{equ:jperp_drift},
\end{align}
where $p_{s\parallel}=\left<(\vect{v}_\parallel-\vect{v}_{s\parallel})
\cdot(\vect{p}_\parallel-\vect{p}_{s\parallel}/n_s)\right>$ and
$p_{s\perp}=0.5\left<(\vect{v}_\perp-\vect{v}_{s\perp})
\cdot(\vect{p}_\perp-\vect{p}_{s\perp}/n_s)\right>$ are parallel and perpendicular
pressures w.r.t the local magnetic field, respectively, $\rho_s$ is the charge density,
$n_s$ is particle number density, $m_s$ is particle mass, and
$d/dt = \partial_t + \vect{v}_s\cdot\nabla$. In the language of particle drifts,
the plasma energization is then associated with parallel electric field, curvature drift,
gradient drift, magnetization, and flow inertia.
\citet{Li2018Roles} reorganized $\vect{j}_{s\perp}\cdot\vect{E}_\perp$ as
\begin{align}
  \vect{j}_{s\perp}\cdot\vect{E}_\perp & =
  \nabla\cdot(p_{s\perp} \vect{v}_E) - p_s\nabla\cdot\vect{v}_E
  - (p_{s\parallel}-p_{s\perp})b_ib_j\sigma_{ij} +
  n_sm_s\frac{d\vect{u}_s}{dt}\cdot{\vect{v}_E} \label{equ:comp_shear},
\end{align}
where $\vect{v}_E$ is the $\vect{E} \times \vect{B}$ drift velocity,
$\sigma_{ij} = 0.5 (\partial_i v_{Ej} + \partial_j v_{Ei} -
(2\nabla\cdot\vect{v}_E\delta_{ij})/3)$ is the shear tensor, and $p_s\equiv(p_{s\parallel} + 2p_{s\perp})/3$ is the effective
scalar pressure. Then, plasma energization is associated with parallel electric field,
flow compression, flow shear, and flow inertia.

\section{Drift description of particle acceleration}
Gyrophase-averaged particle guiding center velocity is
\citep{Northrop1963Adabatic, Webb2009Drift, LeRoux2009Time, LeRoux2015Kinetic}
\begin{align}
  \left<\vect{v}_g\right>_\phi = v_\parallel\vect{b} + \vect{v}_E +
  \frac{M}{q_s}\frac{\vect{B}\times\nabla B}{B^2} +
  \frac{m_sv_\parallel}{q_sB}\vect{b}\times\frac{d\vect{b}}{dt} +
  \frac{M}{q_s}(\nabla\times\vect{b})_\parallel +
  \frac{m_s}{q_sB}\vect{b}\times\frac{d\vect{v}_E}{dt}
\end{align}
where $d/dt=\partial/\partial t + (v_\parallel\vect{b} + \vect{v}_E)\cdot\nabla$,
$M=m_s(\vect{v}_\perp - \vect{v_E})^2/2B$ is particle magnetic moment in the plasma
frame where $\vect{v}_E=0$. The terms on the right are the parallel guiding-center velocity,
$\vect{E}\times\vect{B}$ drift, gradient drift, inertial drift (including curvature drift),
parallel drift, and polarization drift. Assuming the perpendicular flow velocity
$\vect{v}_{s\perp}\approx\vect{v}_E$ and particles are non-relativistic
($\vect{v}\approx\vect{p}/m_s$), the current density associated with particle gradient drift is
\begin{align}
  \int q_s \frac{M}{q_s}\frac{\vect{B}\times\nabla B}{B^2} fd^3p =
  p_{s\perp}\frac{\vect{B}\times\nabla B}{B^3}.
  \label{equ:grad}
\end{align}
The current density associated with particle inertial drift is
\begin{align}
  & \int q_s \frac{m_sv_\parallel}{q_sB}\vect{b}\times\frac{d\vect{b}}{dt} fd^3p \nonumber \\
  = & \int \frac{m_sv_\parallel}{B}\vect{b}\times\left(\frac{\partial \vect{b}}{\partial t} +
  (v_\parallel\vect{b} + \vect{v}_E)\cdot\nabla\vect{b}\right) fd^3p \nonumber \\
  = & \frac{p_{s\parallel} + n_sm_sv_{s\parallel}^2}{B}\vect{b}\times(\vect{b}\cdot\nabla)\vect{b} +
  n_sm_s\frac{v_{s\parallel}}{B}\vect{b}\times
  \left(\frac{\partial\vect{b}}{\partial t} + (\vect{v}_E\cdot\nabla)\vect{b}\right) \nonumber \\
  = & \frac{p_{s\parallel}}{B}\vect{b}\times(\vect{b}\cdot\nabla)\vect{b} +
  \frac{n_sm_s}{B}\vect{b}\times(\vect{v}_{s\parallel}\cdot\nabla)\vect{v}_{s\parallel} +
  \frac{n_sm_s}{B}\vect{b}\times
  \left(\frac{\partial}{\partial t} + \vect{v}_E\cdot\nabla\right)\vect{v}_{s\parallel} \nonumber \\
  = & \frac{p_{s\parallel}}{B}\vect{b}\times(\vect{b}\cdot\nabla)\vect{b} +
  \frac{n_sm_s}{B}\vect{b}\times\left(\frac{\partial}{\partial t} +
  \vect{v}_s\cdot\nabla\right)\vect{v}_{s\parallel},
  \label{equ:inertial}
\end{align}
where we get the current density associated with curvature drift and the flow inertial effect
associated with the parallel component of the flow velocity. The current density
associated with particle parallel drift is
\begin{align}
  & \int q_s \frac{M}{q_s}(\nabla\times\vect{b})_\parallel fd^3p \nonumber \\
  = & \frac{p_{s\perp}}{B}(\nabla\times\vect{b})\cdot\vect{b}\vect{b} \nonumber \\
  = & \left(\nabla\times\frac{p_{s\perp}\vect{B}}{B^2} -
  \nabla\frac{p_{s\perp}}{B}\times\vect{b}\right)\cdot\vect{b}\vect{b}\nonumber \\
  = &  - \left[\nabla\times\frac{p_{s\perp}\vect{B}}{B^2}\right]_\perp +
  \nabla\times\frac{p_{s\perp}\vect{B}}{B^2},
  \label{equ:parallel}
\end{align}
where the first term is the current density associated with magnetization, and
the dot product of the second term with $\vect{E}$ gives
\begin{align}
  & \left(\nabla\times\frac{p_{s\perp}\vect{B}}{B^2}\right)\cdot\vect{E} \nonumber \\
  = & -\nabla\cdot(p_{s\perp}\vect{v}_E) + \frac{p_{s\perp}\vect{B}}{B^2}\cdot
  \nabla\times\vect{E} \nonumber \\
  = & -\nabla\cdot(p_{s\perp}\vect{v}_E) - \frac{p_{s\perp}}{B}\frac{\partial B}{\partial t}
  \label{equ:bmoment}
\end{align}
where we used the Maxwell-Faraday equation, the first term cancels the first term on
the right in Equation~\ref{equ:comp_shear}, and the second term cancels betatron acceleration.
Finally, the energization associated with particle polarization drift is
\begin{align}
  \int q \frac{m_s}{qB}\vect{b}\times\frac{d\vect{v}_E}{dt} fd^3p
  & = \frac{n_sm_s}{B}\vect{b}\times\left(\frac{\partial}{\partial t} +
  \vect{v_s}\cdot\nabla\right)\vect{v}_E,
  \label{equ:polar}
\end{align}
which contributes to the flow inertial term. Combining Equation~\ref{equ:grad}
to~\ref{equ:polar}, we can reproduce Equation~\ref{equ:jperp_drift}.
The total plasma energization is
\begin{align}
  \int q_s\left<\vect{v}_g\right>_\phi\cdot\vect{E} fd^3p +
  \frac{p_{s\perp}}{B}\frac{\partial B}{\partial t}& =
  - p_s\nabla\cdot\vect{v}_E - (p_{s\parallel}-p_{s\perp})b_ib_j\sigma_{ij} +
  n_sm_s\frac{d\vect{u}_s}{dt}\cdot{\vect{v}_E} \label{equ:total_ene},
\end{align}
which is different from Equation~\ref{equ:comp_shear} because of the terms in
Equation~\ref{equ:bmoment}. Table~\ref{tbl:fluid_particle} compares the
two descriptions.
\begin{deluxetable}{r|l}
  \tablecaption{Comparing particle description and fluid description
    of the energization processes\label{tbl:fluid_particle}}
  \tablecolumns{2}
  \tablehead{
  \colhead{Particle description} &
  \colhead{Fluid description (Equ.~\ref{equ:jperp_drift})}
  }
  \startdata
  inertial drift (Equ.~\ref{equ:inertial}) & curvature drift + part of flow inertial term \\
  curvature drift (part of inertial drift Equ.~\ref{equ:inertial}) & curvature drift \\
  gradient drift (Equ.~\ref{equ:grad}) & gradient drift \\
  parallel drift + betatron acceleration (Equ.~\ref{equ:parallel} and~\ref{equ:bmoment}) & magnetization \\
  polarization drift (Equ:~\ref{equ:polar}) & part of flow inertial term \\
  parallel guiding-center velocity & parallel flow velocity \\
  $\vect{E}\times\vect{B}$ drift & $\vect{E}\times\vect{B}$ drift \\
  \enddata
\end{deluxetable}

\acknowledgments
This work was supported by NASA grant NNH16AC60I.
HL and FG acknowledgess the support by DOE/OFES.
We also acknowledge support by the DOE through the LDRD program at LANL.
We gratefully acknowledge our discussions with Bill Daughton, Ari Le, Adam Stanier, and Patrick Kilian.
Simulations were performed with LANL institutional computing.

\bibliography{references}{}
\bibliographystyle{aasjournal}
\end{document}